\documentclass[journal]{IEEEtran}
\ifCLASSINFOpdf
\else
\fi

\usepackage[font=small]{caption}
\usepackage{thmtools,thm-restate}
\usepackage{stfloats}
\usepackage{neuralnetwork}
\usepackage{xpatch}
\makeatletter
\xpatchcmd{\linklayers}{\nn@lastnode}{\lastnode}{}{}
\xpatchcmd{\linklayers}{\nn@thisnode}{\thisnode}{}{}
\makeatother

\usepackage{tikz-cd}
\usepackage{pgfplots}
\usepackage {graphics}
\usepackage {graphicx}\usepackage{subfigure}
\usepackage{titlesec}
\usepackage{float}
\usepackage{indentfirst,latexsym,bm,color}
\usepackage{amsmath,amssymb,amsfonts,amsthm}
\usepackage{CJK}
\usepackage{mathrsfs}
\usepackage{multirow}
\usepackage{extarrows}
\usepackage{booktabs}
\usepackage{makecell}
\usepackage{caption}
\usepackage{algorithm} 
\usepackage{algpseudocode} 
\usepackage[algo2e,ruled, vlined, linesnumbered]{algorithm2e}
\usepackage{tabularx}
\usetikzlibrary{positioning,arrows.meta}
\tikzset{
     block/.style={rectangle, draw, fill=red!40, text width=6em,
                   text centered, rounded corners, minimum height=3em},
     arrow/.style={-{Stealth[]}}
}

\newtheorem{theorem}{Theorem}

\newtheorem{corollary}{Corollary}

\makeatletter
\newcommand{\algmargin}{\the\ALG@thistlm}
\makeatother
\newlength{\whilewidth}
\algdef{SE}[parWHILE]{parWhile}{EndparWhile}[1]
  {\parbox[t]{\dimexpr\linewidth-\algmargin}{%
     \hangindent\whilewidth\strut\algorithmicwhile\ #1\ \algorithmicdo\strut}}{\algorithmicend\ \algorithmicwhile}%
\algnewcommand{\parState}[1]{\State%
  \parbox[t]{\dimexpr\linewidth-\algmargin}{\strut #1\strut}}

\begin{document}
%
\title{Real-time privacy preserving disease diagnosis using ECG signal}

\author{Guanhong~Miao,
        A.~Adam~Ding,
        and~Samuel~S.~Wu  
\thanks{G. Miao and S. Wu are with University of Florida, Gainesville, FL, 32611, USA. e-mail: gmiao@ufl.edu, samwu@biostat.ufl.edu.}
\thanks{A. Ding is with Northeastern University, Boston, MA, 02115, USA. e-mail: a.ding@neu.edu.}
}

\maketitle


\begin{abstract}
The rapid development in Internet of Medical Things (IoMT) boosts the opportunity for real-time health monitoring using various data types such as electroencephalography (EEG) and electrocardiography (ECG). Security issues have significantly impeded the e-healthcare system implementation. Three important challenges for privacy preserving system need to be addressed: accurate diagnosis, privacy protection without compromising accuracy, and computation efficiency. It is essential to guarantee prediction accuracy since disease diagnosis is strongly related to health and life. By implementing matrix encryption method, we propose a real-time disease diagnosis scheme using support vector machine (SVM). A biomedical signal provided by the client is diagnosed such that the server does not get any information about the signal as well as the final result of the diagnosis while the proposed scheme also achieves confidentiality of the SVM classifier and the server's medical data. The proposed scheme has no accuracy degradation. Experiments on real-world data illustrate the high efficiency of the proposed scheme. It takes less than 1 second to derive the disease diagnosis result using a device with 4Gb RAMs, suggesting the feasibility to implement real-time privacy preserving health monitoring. 
\end{abstract}



\section{Introduction}
With the rapid development of data collection via Internet of Medical Things (IoMT), machine learning becomes prevalent in analyzing big data and plays a crucial role in online medical diagnosis. Due to the shortage of experts and high cost in manual diagnosis, machine learning improves the quality of healthcare service and avoids expensive diagnosis expenses. With the emergence of personal device applications, various disease prediction systems have been investigated. Existing works adopt wearable IoT in healthcare to diagnose certain diseases including Alzheimer disease \cite{application_1} and Parkinson's disease \cite{application_2}. 

It is important to develop a high-performance model for real-time and reliable medical diagnosis. Support vector machine (SVM), the state-of-the-art machine learning models, have been investigated for disease prediction systems \cite{application_3,application_4}. Personalized healthcare is benefited from the wearable Internet of Things (IoT) devices. Electroencephalography (EEG) and electrocardiography (ECG) sensors are designed to monitor brain-electrical activity and heart activities, respectively \cite{iot_2018}. SVM has also been studied to predict diseases such as epileptic seizure and arrhythmias by analyzing EEG and ECG activities \cite{signal_2010,signal_2012,signal_2013,EEG_2018,EEG_2018_1}. 

Machine learning-based diagnosis and prediction has also been accompanied by privacy concerns. Health data is considerably sensitive as it contains patient characteristics. For instance, current research shows that EEG can be used to predict viewed images, monitor sleep and reveal user information such as age, gender, and user's illnesses or additions \cite{EEG}. It is essential to protect medical information when performing disease prediction. Moreover, the diagnosis result is also sensitive and should be protected. As an example, patients with epilepsy may experience feelings of shame and isolation \cite{application_5}. Releasing diagnosis result may lead to potential psychological problems.  


In this paper, we focus on real-time health monitoring which should work to detect health emergency timely. Suppose that patient Alice has chronic disease and IoMT device for ECG monitoring is equipped to collect signals frequently, especially during exercise for timely emergency detection. The ECG signal monitoring is especially important to detect chronic heart diseases. We design privacy preserving disease prediction scheme which allows the user to check health status regularly without going to an analysis center or hospital while protecting personal signal record as well as diagnosis result. In the simplest case, the above scenario contains only two parties. One party, referred to as the client, owns a signal generated by the personal device. Another party, referred to as the server, owns the medical data and is supposed to provide disease prediction service. The server can be a single healthcare provider owning one medical database or a cloud platform with access to multiple medical databases. Because the model training is usually time consuming and restricts real-time applications, it is reasonable to build prediction model prior to the request for real-time disease prediction. We assume model training is performed by the server locally prior to the proposed real-time privacy preserving disease diagnosis scheme. 

Various privacy preserving disease diagnosis systems have been proposed for non-ECG data \cite{disease_2019,disease_2018,disease_2019_1,related_01,disease_2020}. Most of the existing secure systems assume that multiple data providers cooperate to build a global model and cannot achieve real-time diagnosis. Different from the general health care data, ECG signals are usually pre-processed by dimension reduction methods \cite{signal_2010,signal_2012,signal_2013,EEG_2018,EEG_2018_1}. So the existing secure systems may not achieve high computation efficiency and accuracy when analyzing ECG data. 

We investigate a real-time privacy preserving disease diagnosis system for ECG data. In the proposed system, the client can achieve privacy preserving medical diagnosis using personal devices. To save time, the disease diagnosis model is built by the server before client request. The server provides disease diagnosis service without leakage of the disease diagnosis model and medical database. In particular, we consider an SVM to classify the disease status for ECG signals collected by the client's personal device. Our privacy preserving system protects data confidentiality using random Gaussian matrix which has been used for signal data encryption previously \cite{prove_0,CPA_intro,CPA_intro1}. Our contributions are as follows.

\begin{enumerate}
    \item We propose a real-time privacy preserving scheme for disease diagnosis for ECG signal data. The proposed scheme protects confidentiality of the client's signal record as well as the diagnosis result. In addition, the server keeps the diagnosis model and medical database secret from the client. 
    \item The secure scheme does not compromise the accuracy. 
    \item Real-world ECG data analysis illustrates that the proposed scheme has high efficiency and accuracy to achieve real-time health monitoring. 
\end{enumerate}

The rest of the paper is organized as follows. Section 2 reviews the related work and Section 3 outlines preliminaries. Section 4 describes the system model and design goal snd Section 5 provides the construction of proposed method. Security analysis is given in Section 6. Section 7 provides the performance evaluations. Finally, Section 8 concludes the paper.

\section{Related work}

With the advanced technology development, ECG signals have been collected for health monitoring such as real-time heart disease prediction \cite{iot_2018,realtime}. Clinically, ECG signals are the most commonly preferred diagnostic tool due to its non-invasiveness and low cost \cite{ECG_intro}. The ECG signal provides vital information regarding the function and rhythm of the heart. Principal component analysis (PCA) is the common technique to process ECG signals and extract features for subsequent analysis \cite{PCA_ECG,signal_2013}. After the signal processing, support vector machine (SVM) classifier has been widely applied to further conduct ECG classification \cite{SVM_ECG,SVM_ECG2,SVM_ECG3}. 

Several cloud-based monitoring systems for ECG data have been investigated previously. A privacy preserving system was proposed for ECG classification assuming that the classification model has been trained and stored in the service provider before service request \cite{homomorphicE}. An ECG analysis algorithm with the implement on an Internet of Things (IoT)-based embedded platform was proposed for real-time monitoring of patients' status \cite{main_3}. This algorithm does not consider the patient privacy. A federated learning-based distributed algorithm provided a solution to build disease diagnosis model using ECG databases owned by different medical institutions with privacy protection \cite{main_2}. Another federated learning-based ECG analysis model was proposed to enable distributed learning among multiple data providers without the need for direct ECG data exchange with the cloud \cite{main_1}.      

Privacy preserving disease diagnosis schemes have been proposed based on various machine learning models. Using homomorphic encryption (HE), a privacy preserving outsourced disease predictor on random forest was proposed to analyze data distributed across different institutions \cite{disease_2019}. Utilizing single-layer perceptron, an efficient and secure disease prediction scheme was investigated using random invertible matrix for data encryption \cite{disease_2018}. Based on Paillier Cryptosystem with
Threshold Decryption (PCTD), a framework for hybrid privacy-preserving clinical decision support system in fog–cloud computing was designed using multiple-layer neural network \cite{related_03}. With homomorphic cryptographic algorithm, a disease risk prediction system including model training and remote disease prediction were proposed \cite{disease_2019_1}. Another privacy preserving disease risk assessment scheme was investigated for multi-outsourced vertical datasets using Paillier cryptosystem \cite{related_01}. Based on multi-label $k$-nearest-neighbors, a secure multi-level medical pre-diagnosis system was proposed with Boneh–Lynn–Shacham (BLS) short signature \cite{related_02}. Using two-trapdoor public-key cryptosystem, a privacy-preserving medical diagnosis mechanism based on extreme gradient boosting (XGBoost) was designed in edge computing \cite{disease_2020}.

Besides the encryption methods as mentioned above, matrix encryption method has been studied for efficient privacy preserving data analysis. For example, an encryption approach was proposed using random orthogonal matrix in \cite{perturbation4}. Data privacy could also be protected via random projection perturbation using dimension reduction approach \cite{multiplicative4}. Elementary matrix transformation was proposed to achieve secure outsourcing face recognition \cite{private_2020}. Secure algorithms for outsourcing matrix operations and outsourcing linear equations were investigated based on matrix encryption \cite{matrix_1,matrix_0}. Privacy preserving outsourced computation was studied using sparse matrix for data encryption \cite{smm}. The comprehensive security analysis was provided for signal data encryption using random Gaussian matrix recently \cite{prove_0,CPA_intro,CPA_intro1}. 

Real-time disease diagnosis can provide a powerful tool to monitor health condition and prevent the worst scenario. While most existing secure systems cannot achieve real-time diagnosis, we noticed a real-time privacy preserving scheme designed for non-ECG data \cite{svm_2016}. The scheme in \cite{svm_2016} assumes that classification model is built by the service provider before client request. Similar assumption has been used in another privacy preserving scheme designed for ECG classification \cite{homomorphicE}. However, the secure scheme in \cite{homomorphicE} cannot achieve real-time classification. Since the real-time disease diagnosis system is lacking for ECG data, our goal is to design such scheme for health monitoring. Particularly, we assume the disease diagnosis model is learned using the ECG data stored in the server prior to the diagnosis request. The proposed scheme can protect the medical database, the client's ECG signals as well as the diagnosis model and results.

\section{Preliminary}
\subsection{ECG signal processing}
The ECG signals are first pre-processed to eliminate the noise using methods such as discrete wavelet transform (DWT) \cite{signal_2010,signal_2012,signal_2013}. In the following post-processing phase, dimension reduction is usually applied to project ECG signals to a low feature space and classification method is then applied for diagnosis \cite{signal_2010,signal_2012,signal_2013,EEG_2018,EEG_2018_1}. 

Consider a disease diagnosis system using ECG signals. Assume the client collects personal ECG signals and submits a disease diagnosis request to the server who owns the diagnosis model and provides the diagnosis service. The general ECG signal processing procedures are summarized as follows.

\begin{enumerate}
    \item Pre-processing (e.g., DWT) is conducted locally in the server for the medical database and the client for personal ECG signals.
    \item In the post-processing procedure, each ECG signal is projected to a new feature space by dimension reduction method (e.g., principal component analysis (PCA)). The server conducts post-processing procedure locally. Since the client does not have access to the dimension reduction model, post-processing procedure for personal ECG signal will be conducted by interacting with the server.
    \item Disease diagnosis is conducted for ECG signals in the new feature space. 
\end{enumerate}


\subsection{Feature extraction}
Principal Component Analysis (PCA) chooses a linear projection that maximizes the variability of all projected samples. Suppose there are $n$ signals in the ECG database, $x_i\in \mathbb{R}^q$ ($i=1,2,\cdots,n$) and $X=[x_1,\cdots,x_n]$. Matrix $X$ contains the training database such that each column corresponds to the signal of one individual. Let $S_T$ be the total covariance matrix and PCA projection matrix is defined as
\begin{displaymath}
W_{pca}=\text{arg }\underset{w}{\text{max}}|W^TS_TW|=[w_1,\cdots,w_p].
\end{displaymath}
The new feature vectors $z_i\in \mathbb{R}^{p}$ are defined by the following linear transformation:
\begin{displaymath}
z_i=W_{pca}^T x_i ~~~~~~~i=1,2,\cdots,n
\end{displaymath}
where $W_{pca}\in \mathbb{R}^{q\times p}$ is a matrix with orthonormal columns.

Other feature extraction algorithms are described in Appendix \ref{appendix_LDA ICA}.

\subsection{Commutative matrix}
Matrix $A$ and $B$ are commutative if $AB=BA$. To maintain high accuracy, the proposed privacy preserving scheme generates commutative matrix for data encryption. We construct commutative encryption matrix based on matrix polynomial (i.e., a polynomial with matrices as variables) \cite{commutative}. Specifically, the server and the client first generate common encryption key $B_0$ (the random invertible matrix). The client then generates a vector of random coefficients $(b_{11}, b_{12}, \cdots, b_{1q})$ and the encryption matrix $B_{11}=\underset{j=1}{\overset{q}{\sum}}b_{1j}B_0^j=b_{11}B_0+b_{12}B_0B_0+b_{13}B_0B_0B_0+\cdots+b_{1q}B^q_0$. Similarly, the server generates a vector of random coefficients $(b_{21}, b_{22}, \cdots, b_{2q})$ and the encryption matrix $B_{22}=\underset{j=1}{\overset{q}{\sum}}b_{2j}B_0^j=b_{21}B_0+b_{22}B_0B_0+b_{23}B_0B_0B_0+\cdots+b_{2q}B^q_0$. Because $B_{11}$ and $B_{22}$ are matrix polynomials of the same matrix $B_0$, it is easy to verify that $B_{11}B_{22}=B_{22}B_{11}$, i.e., $B_{11}$ and $B_{22}$ are commutative.

\section{System model and design goal}
Privacy preserving disease prediction systems have been widely studied. The general IoMT system model comprises three basic parties, namely: data providers, data processing center/server (e.g., hospital) and service requester/client \cite{SVM_2021,disease_2019,disease_2019_1,disease_2019_2}. The server (i.e., the data processing center) trains diagnosis/prediction model using health records provided by data providers. For each service request, the server performs prediction phase using model parameters derived from the training phase. In the model training phase, encryption algorithm brings additional computation cost and cloud service provider is involved in some secure IoMT systems for efficiency purpose. To achieve real-time health monitoring and disease prediction, we assume model training is done before service request  \cite{svm_2016}. Since a single record from one individual should not have significant effect on the robust model, it is reasonable to update model on the backend after the server receives a batch of new records from authorized participants. 

\subsection{System model}
We focus on how the server offers privacy-preserving and efficient medical diagnosis service to the client whose medical data is sensitive. The proposed scheme involves two parties: the client who has personal ECG signals and the server which has access to the medical database and provides the service of disease diagnosis. The client wants to know the disease risk with ECG signal obtained by personal devices with privacy protection. If information of the client is disclosed, the insurance companies can restrain the client from coverage and the sensitive information leakage can also lead to some consequences such as public humiliation and losing jobs. On the other hand, the server also considers the diagnosis model and medical database as private information and is not willing to reveal. 

\begin{figure}[h]
\centering
\includegraphics[scale=0.32]{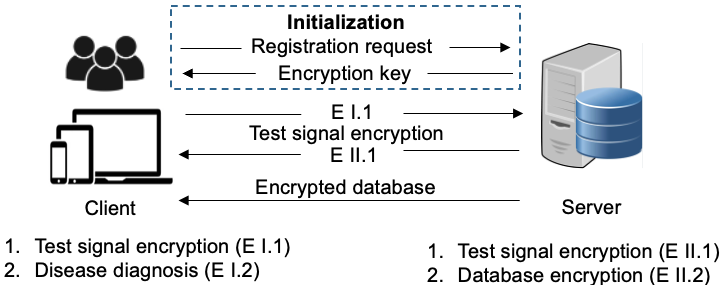}
\caption{The overview of privacy preserving disease diagnosis framework}\label{fig1}
\end{figure}

We consider the server owns an SVM classifier built upon existing medical database and provides medical diagnosis service for registered clients. Before disease diagnosis using SVM, the server projects the medical database to a new feature space (i.e., the post-processing procedure for ECG signals). Specifically, each ECG signal, a $q$-dimensional vector, is mapped to a $p$-dimensional feature space, where $p<q$. The new feature vector $z_i\in \mathbb{R}^{p}$ is defined by the following linear transformation:
\begin{equation}\label{equation1}
z_i=W x_i ~~~~~~~i=1,2,\cdots,n
\end{equation}
where $W\in \mathbb{R}^{p\times q}$ is the projection matrix. For PCA, $W=W_{pca}^T$.  

Disease diagnosis is performed in a privacy preserving way such that data transmitted between the client and the server is encrypted. Each party conducts two procedures to achieve privacy preserving disease diagnosis as presented in Figure \ref{fig1}. The client encrypts ECG signal $t$ and transmits to the server (E I.1); the server further encrypts received signal to get $t^{*}$ and transmits it to the client (E II.1); the server maps the whole database to a new feature space with the projection $W$ and then encrypts both new database and projection $W$ and transmits to the client (E II.2); the client projects $t^{*}$ to the new feature space using encrypted $W$ and then performs disease prediction using encrypted database (E I.2).

\subsection{Threat model}
We follow the threat model discussed in previous privacy preserving medical diagnosis framework \cite{svm_2016}. In our medical scenario, we consider the server and client are semi-honest (i.e., follow the scheme and may be curious about the intermediate data). Specifically, the server provides the disease diagnosis correctly, but is curious to the client's medical information. The client honestly conducts the operations to derive the diagnosis result, but also tries to decrypt the parameters of the diagnosis model and the medical database in the server. Moreover, the client may try to access the diagnosis service without registering. To guarantee the privacy of the client's medical information and the confidentiality of diagnosis model and medical database, the following security requirements should be satisfied. 

\begin{enumerate}
    \item {\bf Privacy.} Protecting the client's medical information from the server, i.e., the server cannot recover the client's ECG signal in plaintext form. Moreover, the server cannot obtain the disease diagnosis result.
    \item {\bf Confidentiality.} Keeping the diagnosis model and medical database secret from the client, i.e., the client cannot identify the parameters of the diagnosis model or medical database from the encrypted database.
    \item {\bf Authentication.} Authenticating an encrypted ECG signal that is really sent by a legal client, i.e., if an illegal client forges a data query, this malicious operation should be detected. 
\end{enumerate}

In the proposed scheme, matrix encryption technique is applied to achieve privacy protection and we discuss three major types of attack models (i.e., ciphertext only attack, known plaintext attack, chosen plaintext attack) investigated in previous matrix encryption systems \cite{CPA_intro1,CPA_intro2,CPA_intro3}. Among these 3 attack models, chosen plaintext attack is more powerful and more threatening than the other 2 attack models \cite{CPA_intro1,CPA_intro2,CPA_intro3}. In the chosen plaintext attack (CPA), it is presumed that the pair of the chosen input plaintexts (i.e., the original data) and the corresponding cyphertexts (i.e., the encrypted data) are accessible to the adversary. Specifically, the client generates fake ECG signals and sends to the server for encryption. After receiving the encrypted data, the client conducts the attack to recover the encryption matrix and the original ECG database owned by the server. We aim to design a secure privacy preserving scheme resilient to CPA.

\subsection{Design goals}
The design goals are summarized as follows.
\begin{itemize}
      \item Privacy: The proposed privacy preserving scheme is secure to protect data confidentiality and achieves authentication for legal clients.
      \item Correctness: The proposed scheme is able to provide the same result as that of the scheme using SVM in the plain domain.
      \item Efficiency: Considering the real-time requirements of disease diagnosis, the scheme should have low overhead in terms of computation and communication.
\end{itemize}





\section{The proposed method}


\subsection{System initialization}
During the registration, the client sends the personal information to the server for identification. The server then generates a unique encryption key $B_0$, a $q\times q$ random invertible matrix with $q$ unique eigenvalues and sends the key to the client. Each registered client receives a unique encryption key.

\subsection{Data encryption} 

\begin{algorithm2e}
\SetKwBlock{kwI}{Client}{end}
\SetKwBlock{kwII}{Server}{end}
\SetKwBlock{reg}{Registration}{end}
\reg{ 
The client sends his/her personal information to the server for identification\;
The server first verifies the client's information. Then the server generates a unique encryption key $B_0$ (a $q\times q$ random invertible matrix with each element following normal distribution $N(0,\sigma^2)$. We require $B_0$ has $q$ unique eigenvalues. $B_0$ is regenerated if it has $<q$ unique eigenvalues.) and sends the key to the client\;
}
\kwI{ 
Generate a random coefficient vector $(b_{11}, b_{12}, \cdots, b_{1q})$, $B_{11}=\underset{j=1}{\overset{q}{\sum}}b_{1j}B_0^j$ and compute $B_{11}t$\;
Generate a random $q$-dimensional vector $t_v$ such that the summation of the $q$ elements equals 1; then compute $t^*_v=B_0t_v$\; 
Normalize ECG signal $t$ and then encrypt it using $B_{11}$. Send the encrypted $B_{11}t$ and $t^*_v$ to the server\;
}
\kwII{
         Compute $\hat{t}^*_v=(B_0)^{-1}t^*_v$ and verify whether the summation of the $q$ elements in $\hat{t}^*_v$ equals 1\;
         \If{the summation of $\hat{t}^*_v$ is 1}
         {
          Generate a random coefficient vector $(b_{21}, b_{22}, \cdots, b_{2q})$ and computes $B_{22}=\underset{j=1}{\overset{q}{\sum}}b_{2j}B_0^j$\;
	 Generate a $p\times p$ random invertible matrix $A$ with each entry following $N(0,\sigma^2_A)$ and let $A_I=(AA^T)^{-1}$\;
	Compute $W^{*}=AWB_{22}$ and $X_{W}^{*}=AWX$\;
	Compute $t^*=B^{-1}_{22}B_{11}t$ and sends to the client\;
        Send $W^{*}$, $X_{W}^{*}$ and $A_I$ to the client\;
         }
}
\kwI{
Compute $t_{W}^{*}=W^{*}B_{11}^{-1}t^{*}$\;
}
\caption{Data encryption}\label{alg1}
\end{algorithm2e}

{\bf Authentication.} The client generates a random $q$-dimensional vector $t_v$ such that the summation of the $q$ elements equals 1. Then the client encrypts $t_v$ using the encryption key $B_0$ and sends the encrypted vector $t^*_v=B_0t_v$ to the server. Once receiving the diagnosis request, the server first verifies whether the client is already registered. Specifically, the server computes $\hat{t}^*_v=(B_0)^{-1}t^*_v$ and checks whether the $q$ elements in $\hat{t}^*_v$ equals 1.

{\bf Client signal encryption.} In order to encrypt signal $t$ while maintaining data utility, the client and the server generate commutative encryption matrices ($B_{11}$ and $B_{22}$) using the same encryption key $B_0$. Specifically, the client generates a random coefficient vector $(b_{11}, b_{12}, \cdots, b_{1q})$ and $B_{11}=\underset{j=1}{\overset{q}{\sum}}b_{1j}B_0^j$. Similarly, the server generates random coefficient vector $(b_{21}, b_{22}, \cdots, b_{2s_0})$ and $B_{22}=\underset{j=1}{\overset{s_0}{\sum}}b_{2j}B_0^j$. The client encrypts $t$ using $B_{11}$ and then sends to the server. After receiving $B_{11}t$, the server calculates $t^*=B^{-1}_{22}B_{11}t$ and sends back to the client. The client can send multiple ECG signals to the server for diagnosis. To protect against adversary attack, the number of signals sent to the server at each diagnosis request is limited to be smaller than $q$ (Section \ref{CPA}).

{\bf Medical database encryption.} The server generates random invertible matrix $A$ to encrypt projection $W$ and $WX$. The encrypted data $W^{*}=AWB_{22}$ and $X_{W}^{*}=AWX$ are transmitted to the client. To derive accurate result, the server further computes $A_I=(AA^T)^{-1}$ and sends to the client.

\begin{figure}[htbp]
\begin{small}
\begin{center}
\begin{tikzpicture}[scale=2]
    \tikzstyle{ann} = [draw=none,fill=none,right]
    \matrix[nodes={draw,minimum size=10mm,inner sep=0pt},
        row sep=0.25cm,column sep=0.5cm,every even column/.style={column sep=1.6cm}] {
    \node[draw=none,fill=none] {}; &
    \node[draw=none,fill=none] {Client}; &
    \node[draw=none,fill=none] {Server};\\
    \node[draw=none,fill=none] {Step 0}; &
    \node[rectangle] {$B_0$, $t$}; &
     \node[rectangle,minimum height=10mm,minimum width=18.5mm] {\begin{tabular}{c} $B_0$, $W$, $X$ \end{tabular}};\\
     \node[draw=none,fill=none] {E I.1}; &
   \node[rectangle] (d)  {$B_{11}t$};  &
     \node[rectangle,minimum height=10mm,minimum width=18.5mm] (c) {$B_{22}$};\\
    \node[draw=none,fill=none] {E II.1}; &
    \node[rectangle] (f)  {$t^{*}$};  &
    \node[rectangle,minimum height=10mm,minimum width=18.5mm] (e) {$t^{*}$};\\
    \node[draw=none,fill=none] {E II.2}; &
    \node[rectangle] (b)  {$t^{*}_W$};  &
     \node[rectangle,minimum height=10mm,minimum width=18.5mm] (a) {\begin{tabular}{c} $A$, $W^{*}$, $X_W^{*}$ \end{tabular}};\\
        \node[draw=none,fill=none] {E I.2}; &
    \node[draw=none,fill=none]  {Prediction};  &
    \node[draw=none,fill=none]   {};\\
    };
     \draw[arrow] (a)  --  (b) node [above,pos=0.5] {$W^{*}$, $X_W^{*}$} node [below,pos=0.5] { $A_I$};
     \draw[arrow] (d)  --  (c) node [above,pos=0.5] {$B_{11}t$};
     \draw[arrow] (e)  --  (f) node [above,pos=0.5] {$t^{*}$};
\end{tikzpicture}
\end{center}
\end{small}
\caption[]{Privacy preserving procedures. In each step, each party generates or has access to matrices in the boxes. Matrices transferred between two parties are above or below the arrows.}\label{fig_1}
\end{figure}
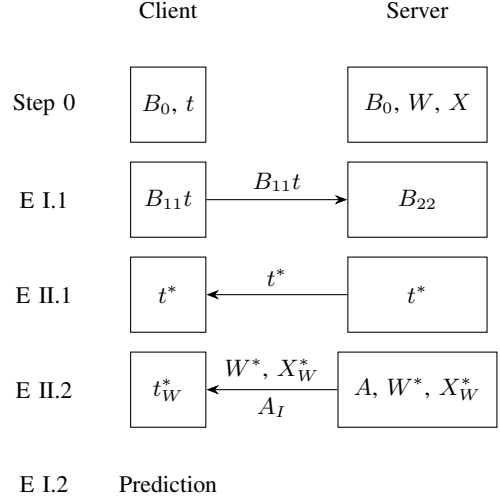

Algorithm \ref{alg1} provides details of the proposed encryption method. Figure \ref{fig_1} depicts data communication details between the client and the server. 

\subsection{Disease prediction} \label{acc}
We design decryption method such that the client derives accurate disease prediction with encrypted data. The privacy preserving disease classification model is based on support vector machine (SVM) and the accuracy is not degraded.

We first define notations for signal $x_i$ in database $X$ and test signal $t$: $x_{iW}=Wx_i$, $x_{iW}^*=Ax_{iW}$, $t_W=Wt$, $t_W^*=At_W$. $x_{iW}^*$ and $t_W^*$ are released to the client while $A_I=(AA^T)^{-1}$ is also sent to the client for decryption.

In the model training stage, the server builds SVM on new feature database $WX$. To predict the client's disease status, the server sends model parameters $\beta_0$ and $\alpha_iy_i$ ($i=1,\cdots,n$) to the client. A SVM classifier \cite{statistics} calculates the classification result of projected test signal $t_W$ using 
\begin{displaymath}
f(t_W)=sgn(\overset{n}{\underset{i=1}{\sum}}\alpha_iy_iK(t_W,x_{iW})+\beta_0)
\end{displaymath} 
where $y_i$ is the class label of the training signal $x_{iW}$. The most common kernel functions are (1) linear kernel function: $K(t_W,x_{iW})=t_W^Tx_{iW}$, (2) polynomial function: $K(t_W,x_{iW})=(t_W^Tx_{iW}+1)^d$ where $d$ is the degree, (3) Gaussian radial basis function: $K(t_W,x_{iW})=exp(-\gamma||t_W-x_{iW}||^2)$ for $\gamma>0$, (4) hyperbolic tangent function: $K(t_W,x_{iW})=tanh(\kappa t_W^Tx_{iW}+c)$ for some $\kappa$ and $c$. 

\textbf{Correctness} Let $f(t_W^{*})$ denote SVM result calculated by the client using encrypted data, i.e., $f(t_W^{*})=\overset{n}{\underset{i=1}{\sum}}\alpha_iy_iK(t_W^{*},x^{*}_{iW})+\beta_0$. Besides $W^*$ and $X^*_W$, the server also sends $A_I=(AA^T)^{-1}$ to the client to ensure no accuracy degradation. It is easy to verify the following two equations.

1. \begin{displaymath}
t_W^{*T}A_Ix_{iW}^*=t_W^TA^TA_IAx_{iW}=t_W^Tx_{iW}.
\end{displaymath}

2. \begin{displaymath}
\sqrt{(x_{iW}^*-t_W^*)^TA_I(x_{iW}^*-t_W^*)}
\end{displaymath}
\begin{displaymath}
=\sqrt{(Ax_{iW}-At_W)^TA_I(Ax_{iW}-At_W)}
\end{displaymath}
\begin{displaymath}
=\sqrt{(x_{iW}-t_W)^T(x_{iW}-t_W)}
\end{displaymath}
\begin{displaymath}
=||x_{iW}-t_W||_2.
\end{displaymath}

These two equations ensure that each kernel function using encrypted data as inputs has identical output as using original data as inputs.

\section{Security analysis}\label{privacy}   
In this section, we prove that the proposed encryption method is secure to protect data confidentiality. In addition, simulations are conducted to illustrate that the encrypted data is distinct from the original data in Section \ref{distance_x_axb}.

\subsection{Security measures} \label{measure_privacy}
Similar to previous privacy preserving schemes using matrix encryption to protect signal data \cite{prove_0,CPA_intro,CPA_intro1}, we consider the indistinguishability \cite{indistinguishability} to formalize the notion of computational security. The cryptosystem is said to have the indistinguishability if no adversary can determine in polynomial time which of the two plaintexts corresponds to the ciphertext, with probability significantly better than that of a random guess \cite{CPA_intro, CPA_intro1}. In other words, in a cryptosystem with indistinguishability, an adversary is unable to learn any partial information of the plaintext in polynomial time from a given ciphertext. 

Let $d_{TV}(p_1,p_2)$ be the total variation (TV) distance \cite{TV_distance} between probability distributions $p_1=P(y|t_1)$ and $p_2=P(y|t_2)$. Then, it is readily checked from \cite{distance_book} that the probability that an adversary can successfully distinguish the plaintexts by a binary hypothesis test is bounded by
\begin{displaymath}
p_d\le \frac{1}{2}+\frac{d_{TV}(p_1,p_2)}{2}
\end{displaymath}
where $d_{TV}(p_1,p_2)\in [0,1]$. Therefore, if $d_{TV}(p_1,p_2)=0$, the probability of success is at most that of a random guess, which leads to the indistinguishability \cite{indistinguishability}. 

Since computing $d_{TV}(p_1,p_2)$ directly is difficult \cite{distance_book1}, we employ an alternative distance metric to bound the TV distance. In particular, the Hellinger distance \cite{TV_distance}, denoted by $d_{H}(p_1,p_2)$, is useful by giving both upper and lower bounds on the TV distance \cite{distance_tv}, i.e.,
\begin{displaymath}
d^2_H(p_1,p_2)\le d_{TV}(p_1,p_2)\le d_{H}(p_1,p_2)\sqrt{2-d^2_{H}(p_1,p_2)}
\end{displaymath}
where $d_{H}(p_1,p_2)\in [0,1]$. Moreover, if the ciphertext $y$ conditioned on $t_h$ is a jointly Gaussian random vector with zero mean and covariance matrix $C_h$, where $h\in \{1,2\}$, the Hellinger distance between the multivariate Gaussian distributions $p_1$ and $p_2$ is given by \cite{extra_1} and \cite{extra_2}
\begin{displaymath}
d_H(p_1,p_2)=\sqrt{1-\frac{|C_1|^{\frac{1}{4}}|C_2|^{\frac{1}{4}}}{|C_3|^{\frac{1}{2}}}}
\end{displaymath}
where $C_3=\frac{C_1+C_2}{2}$. The formal definitions and properties of TV and Hellinger distances are given in \cite{TV_distance,distance_book,distance_book1}.

Mutual information is another metrics to evaluate the distance of 2 distributions \cite{prove_0}. Let $I(x_1;x_2)$ be the mutual information between $x_1$ and $x_2$. We evaluate the privacy protection level of the proposed encryption scheme using TV distance and mutual information as follows.

\subsection{Security of the proposed encryption method} \label{CPA}
Table \ref{tab_dif_2} summarizes the original and transmitted data accessible to the client and server, respectively. $W$ is the projection matrix derived by the server using the entire data $X$. We assume $W$ is secure. 

\begin{table}[h]
\caption[]{Matrices accessible to the client and server in the proposed method. Original datasets are sensitive information. Transmitted data are transferred between the client and server.}\label{tab_dif_2}
\begin{small}
\begin{center}
\begin{tabular}{c c c }
\toprule
Dataset & Client & Server \\
\midrule
Original & $t$ & $X$, $W$ \\  
Transmitted & $B_{11}t$ & $AWX$, $AWB_{22}$, $B_{22}^{-1}B_{11}t_i$ \\
\bottomrule
\end{tabular}
\end{center}
\end{small}
\end{table}

In our scheme, commutative matrices are generated by the client and server to encrypt the private signal and data. Specifically, the client and server generates the same encryption key $B_0$ and then the commutative encryption matrices (e.g., $B_{11}$, $B_{22}$) are generated in the form of its matrix polynomial (Algorithm 1), with the random coefficient vectors $(b_{11},b_{12},\cdots,b_{1q})$ are randomly generated by the client and $(b_{21},b_{22},\cdots,b_{2q})$ are randomly generated by the server independently. As shown in Appendix \ref{appendixA}, $B_{ii}$ is determined by the random coefficients $(b_{i1},b_{i2},\cdots,b_{iq})$ ($i=1,2$) of matrix polynomials and cannot be recovered without access to these coefficients.

We first discuss the security of $B^{-1}_{22}B_{11}t$. The client may perform chosen plaintext attack (CPA) by inserting fake data in the proposed scheme. Algorithm \ref{alg2} summarizes the chosen plaintext attack indistinguishability experiment. The client gets $B_{22}t_h=(\underset{j=1}{\overset{q}{\sum}}b_{2j}B_0^{(j-1)})B_0t_h$ $(h=1\text{ or }2)$. It can be split into 2 encryption functions, i.e., $f_1(t_h)=B_0t_h$ and $f_2(f_1(t_h))=(\underset{j=1}{\overset{q}{\sum}}b_{2j}B_0^{(j-1)})f_1(t_h)$. We first show that the encryption function $f_1$ is indistinguishable and then show $f_2$ is indistinguishable. Let $y_h=B_0t_h$ $(h=1,2)$.

\begin{algorithm2e}
The client generates two ECG signals $t_1$ and $t_2$, encrypts them using encryption matrix $B_{11}$ and submits to the server\;
The server randomly selects a received signal $B_{11}t_i$ ($i=1$ or 2) and further encrypts it using $B_{22}$ to get $t^*=B_{22}B_{11}t_i$. The server sends $t^*$ to the client\;
The client decrypts $B_{11}$ from $t^*$ (i.e., $t^{**}=B^{-1}_{11}t^*=B_{22}t_i$). After the decryption, the client aims to figure out which signal was encrypted\;
Let $j$ be the client's guess. If $j=i$, the output of the experiment is 1. Otherwise, the output is 0\;
\caption{CPA indistinguishability experiment}\label{alg2}
\end{algorithm2e}

\begin{theorem}\label{theorem1.} The worst-case lower and upper bounds on $d_{TV}(p_1,p_2)$ are given by 
\begin{displaymath}
d_{TV,low}=1-(\frac{2||t_1||||t_2||}{||t_1||^2+||t_2||^2})^{q/2},
\end{displaymath}
\begin{displaymath}
d_{TV,up}=\sqrt{1-(\frac{2||t_1||||t_2||}{||t_1||^2+||t_2||^2})^{q}}
\end{displaymath}
where $p_1=P(y|t_1)$ and $p_2=P(y|t_2)$.
\end{theorem}

\begin{proof}
Based on the proof of [\cite{CPA_intro}, Lemma 1], the covariance matrix of $y_h=B_0t_h$ conditioned on the plaintext $t_h$ is $C_h=||t_h||^2I$. Therefore, $C_3=\frac{C_1+C_2}{2}=(\frac{||t_1||^2+||t_2||^2}{2})I$. It is obvious that 
\begin{displaymath}
|C_h|=(||t_h||^2)^q,~~~h\in \{1,2\}
\end{displaymath}
and
\begin{displaymath}
|C_3|=(\frac{||t_1||^2+||t_2||^2}{2})^q.
\end{displaymath}
So 
\begin{displaymath}
d_H(p_1,p_2)=\sqrt{1-\frac{|C_1|^{\frac{1}{4}}|C_2|^{\frac{1}{4}}}{|C_3|^{\frac{1}{2}}}}=\sqrt{1-\frac{2||t_1||||t_2||}{||t_1||^2+||t_2||^2})^{q/2}}.
\end{displaymath}
According to the inequality of the Hellinger distance and the TV distance given in Section \ref{measure_privacy}, it is easy to derive the lower and upper bounds of the TV distance. 
\end{proof}

\begin{theorem}\label{theorem2.} $d_{TV,up}(\tilde{p}_1,\tilde{p}_2)\le d_{TV,up}(p_1,p_2)$ where $\tilde{p}_h$ denotes the probability distribution of the output of $f_2(f_1(t_h))$ and $p_h$ denotes the probability distribution of the output of $f_1(t_h)$ ($h\in \{1,2\}$).
\end{theorem}

\begin{proof}
Hellinger distance can be expressed as a function of R\'{e}nyi divergence \cite{postprocess}, i.e., 
\begin{displaymath}
d_H(p_1,p_2)=\sqrt{2(1-e^{-\frac{1}{2}D_{\frac{1}{2}}(p_1|p_2)})}
\end{displaymath}
where $D_{\frac{1}{2}}(p_1|p_2)$ denote the R\'{e}nyi divergence of $p_1$ from $p_2$. Based on the data processing inequality [\cite{postprocess}, Theorem 1], $D_{\frac{1}{2}}(\tilde{p}_1|\tilde{p}_2)\le D_{\frac{1}{2}}(p_1|p_2)$. So $d_H(\tilde{p}_1,\tilde{p}_2)\le d_H(p_1,p_2)$. Because $0\le d_H(p_1,p_2)\le 1$ and function $u(\sqrt{2-u^2})$ is monotonically increasing for $0\le u\le 1$, 
\begin{displaymath}
d_{TV}(\tilde{p}_1,\tilde{p}_2)\le d_{H}(\tilde{p}_1,\tilde{p}_2)\sqrt{2-d^2_{H}(\tilde{p}_1,\tilde{p}_2)}
\end{displaymath}
\begin{displaymath}
\le d_{H}(p_1,p_2)\sqrt{2-d^2_{H}(p_1,p_2)}.
\end{displaymath}
In other words, $d_{TV,up}(\tilde{p}_1,\tilde{p}_2)\le d_{TV,up}(p_1,p_2)$. 
\end{proof}

\begin{corollary}\label{corollary1.} The success probability of an adversary in the indistinguishability experiment is bounded by 
\begin{displaymath}
p_d\le \frac{1}{2}+\frac{1}{2}\sqrt{1-(\frac{2||t_1||||t_2||}{||t_1||^2+||t_2||^2})^{q}}.
\end{displaymath}
In particular, if each plaintext has constant energy, the cryptosystem has the indistinguishability, since $p_d\le 0.5$ for $||t_1||=||t_2||$.
\end{corollary}

Corollary 1 ensures that no adversary can learn any partial information about the plaintext from a given ciphertext, as long as each plaintext has constant energy. Because each ECG signal is standardized to zero mean and unit variance, the encryption function $f_2(f_1(t))=B_{22}t$ has the indistinguishability. So the CPA indistinguishability experiment (Algorithm \ref{alg2}) has $P(CPA=1)=p_d\le \frac{1}{2}$.

Because $W$ has orthonormal rows (i.e., the energy of each row equals 1), $WB_{22}$ also has the indistinguishability and the adversary cannot learn any partial information about $W$ from $WB_{22}$. So $AWB_{22}$ also has the indistinguishability based on the data processing inequality (Theorem \ref{theorem2.}). 

According to Proposition 1 in \cite{prove_0}, we can also get $I(t;B_{22}t)=I(1;B_{22}t)$ for any standardized ECG signal $t$, indicating that the proposed encryption scheme is perfectly secure \cite{prove_0}.

Next we discuss the security of $AWX$ where $A$ is the encryption matrix in which each element in $A$ is randomly generated from Gaussian distribution $N(0,\sigma^2)$. $W$ is derived by the server and we assume it is secure. Each ECG signal $x$ (i.e., each column in $X$) is standardized before encryption. The encryption $AWX$ does not achieve the indistinguishability since $W$ projects data to a lower feature space and this project doesn't guarantee that columns in $WX$ has constant energy. Alternatively, we use mutual information to evaluate the privacy protection level of $AWX$. Let $x_{W}=Wx$, $x^*_{W}=Ax_{W}$ and $I(x_{W},x^*_{W})$ be the mutual information between $x_{W}$ and $x^*_{W}$. Define $\mathcal{E}_{x_{W}}=||x_{W}||^2_2$. Based on Proposition 1 in \cite{prove_0}, we have $I(x_{W};x^*_{W})=I(\mathcal{E}_{x_{W}};x^*_{W})$. This result says that $AWx$ does not reveal anything more about $Wx$ than its energy, indicating that exposing $\mathcal{E}_{x_{W}}$ does not increase the disclosure risk of the proposed encryption method.

\subsection{Security of model parameters}
SVM parameters $\alpha_i y_i$ ($i=1,\cdots,n$) and $\beta_0$ are sent to the client. $y_i$ has two possible values, -1 or 1. So the probability of an adversary recovering $\alpha_i$ ($i=1,\cdots,n$) is $P(\text{recover SVM parameters})=2^{-n}$ which is a negligible function of $n$ (i.e., No. of samples in the SVM training database $X$). Therefore, the components of the SVM classifier are privacy preserving in the proposed scheme.

\section{Performance analysis} 
We use two ECG heartbeat datasets, PTB Diagnostic ECG data \cite{PTB} and MIT-BIH Arrhythmia data \cite{MIT-BIH,ECG_data}, to evaluate the performance of the proposed scheme. There are a total of 14552 samples in the PTB data with 10506 from individuals who have heart disease and 4046 from healthy individuals. The MIT-BIH data consists of 109446 ECG recordings with 90589 of them correspond to normal heart beats and 18857 correspond to abnormal heart beats.

Each dataset is split into the training set and testing set. Assume the training set is the original data stored in the server. The server builds SVM for disease prediction using the training set. Suppose a set of clients hold ECG signals in the testing set. Each client may hold one or multiple signals. The entire testing set is used to evaluate the prediction accuracy of SVM built by the server. In the PTB data, we randomly select 10000 samples as the training set and the remaining 4552 samples are used as the testing set. In the MIT-BIH data, 87554 samples are randomly selected to be included in the training set and the remaining 21892 samples are included in the testing set. 

ECG signals are preprocessed and segmented following the extraction approach described in \cite{preprocess}. We assume the signal extraction and preprocessing procedures are performed by the local computation in the server and the client. The server uses the preprocessed signal data to build SVM and the client preprocesses the personal signal with the same procedures for disease prediction.

The experimental environment is configured on the University of Florida Hipergator 3.0 (i.e., high-performance computing cluster) with 1 CPU and 4Gb RAMs.

\subsection{The encrypted data is distinct from the original data}\label{distance_x_axb}
{\bf Commutative encryption matrix $B_{11}$ and $B_{22}$.} We evaluate the difference between the original signal $t$ and encrypted signal $B_{11}t$ and $B_{22}t$ where $B_{11}$ and $B_{22}$ are commutative matrices generated using the same encryption key $B_0$. Two matrices, mean squared error (MSE) and mean absolute difference (MAD), are used to measure the distance between $t$ and $B_{11}t$ (i.e., D1), $t$ and $B_{22}t$ (i.e., D2), as well as $B_{11}t$ and $B_{22}t$ (i.e., D3). Specifically, MSE between signal $t_1$ and $t_2$ is defined as $\overset{p}{\underset{i=1}{\sum}}(t_1(i)-t_2(i))^2/p$ and MAD is defined as $\overset{p}{\underset{i=1}{\sum}}|t_1(i)-t_2(i)|/p$ where $t_j(i)$ denotes the $i$-th element in signal $t_j$ ($j=1,2$).

\begin{figure}[h]
\centering
\includegraphics[scale=0.4]{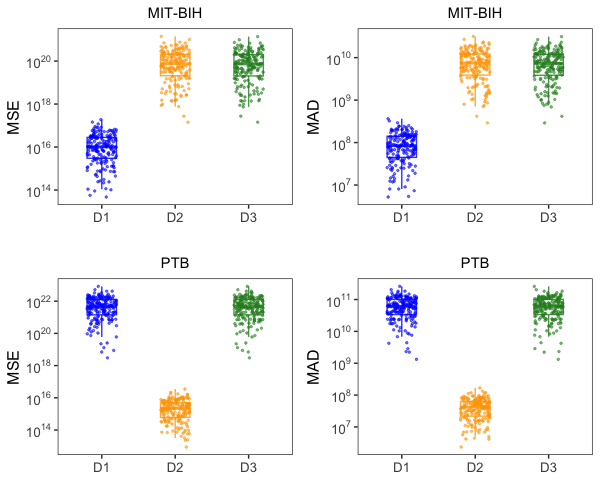}
\caption{Boxplot showing the difference of the original signal ($t$), the encrypted signal I ($B_{11}t$) and II ($B_{22}t$). MSE: mean squared error, MAD: mean absolute difference. D1: distance between $t$ and $B_{11}t$; D2: distance between $t$ and $B_{22}t$; D3: distance between $B_{11}t$ and $B_{22}t$. }\label{fig_p1}
\end{figure}

In the simulation, we first generate the encryption key $B_0$ with each element in the matrix following normal distribution $N(0.1,1)$. The encryption matrices are then set as $B_{11}=3B_0^3+9.1B_0^6$ and $B_{22}=4.3B_0^4+2B_0^8$ for MIT-BIH data encryption and $B_{11}=1.7B_0^4+12B_0^8$ and $B_{22}=6B_0^3+3.4B_0^6$ for PTB data encryption. These 2 commutative encryption matrices are used to encrypt ECG signals in the MIT-BIH and PTB data. MSE and MAD between the original and encrypted signals are calculate for each ECG signal in each dataset. Figure \ref{fig_p1} shows the range of MSE and MAD for the original and corresponding encrypted ECG signals. As shown in the boxplot, although $B_{11}$ and $B_{22}$ are generated using the same encryption key, $B_{11}t$ is significantly different from $B_{22}t$ and both the encrypted signals are different from the original signal $t$. Figure \ref{scatter_p1} (Appendix \ref{appendix_plot}) further shows the difference of the first original signal and the encrypted signal of each ECG dataset.

\begin{figure}[h]
\centering
\includegraphics[scale=0.4]{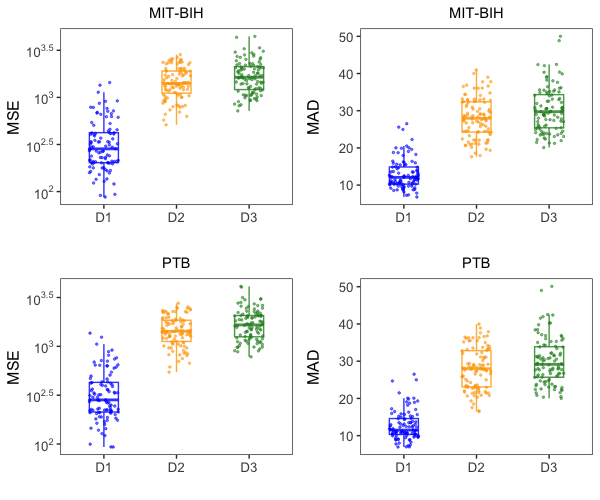}
\caption{Boxplot showing the difference of the original signal in the new feature space ($Wt$) and the encrypted signal I ($A_1Wt$) and II ($A_2Wt$). MSE: mean squared error, MAD: mean absolute difference. D1: distance between $t$ and $B_{11}t$; D2: distance between $t$ and $B_{22}t$; D3: distance between $B_{11}t$ and $B_{22}t$. }\label{fig_p2}
\end{figure}

{\bf Random invertible encryption matrix $A$.} The ECG signal data in the new feature space (i.e., $WX$) is encrypted by the random invertible encryption matrix $A$. To measure the privacy protection provided by the encryption matrix $A$, we calculate MSE and MAD between the original signal in the new feature space ($Wt$) and the encrypted signal ($A_1Wt$, $A_2Wt$ where $A_1$ and $A_2$ are two random invertible matrices). In the simulation, $A_1$ is generated from the exponential distribution and $A_2$ is generated from the uniform distribution. Figure \ref{fig_p2} shows the range of MSE and MAD for the original and corresponding encrypted ECG signals in the new feature space. $A_1Wt$ is significantly different from $A_1Wt$ and both the encrypted signals are different from the original signal $Wt$. Figure \ref{scatter_p2} (Appendix \ref{appendix_plot}) further presents the difference of the first original signal and the encrypted signal of each ECG dataset.

\subsection{Performance evaluation}

\begin{table}[h]
\caption[]{Performance evaluation of PTB Diagnostic ECG data (assume the client request prediction for a total of ten ECG signals)}\label{tab_d1}
\begin{scriptsize}
\begin{center}
\begin{tabular}{c c c c c c}
\toprule
& \multicolumn{5}{c}{No. of principal components}  \\
\cmidrule{2-6} 
 & 10 & 20 & 40 & 60 & 100 \\
\midrule
 \begin{tabular}{@{}c@{}} AUC \\ (proposed scheme) \end{tabular} & 97.27 & 98.41 & 98.35 & 98.44 & 98.04 \\
  \begin{tabular}{@{}c@{}} AUC \\ (non-secure SVM) \end{tabular} & 97.27 & 98.41 & 98.35 & 98.44 & 98.04 \\
Computation & 10 ms & 10 ms & 20 ms & 20 ms & 20 ms \\  
Communication* & 0.4 Mb & 0.8 Mb & 1.6 Mb & 2.3 Mb & 3.9 Mb \\  
\bottomrule
\end{tabular}
\end{center}
~~~~~~* 32-bit precision.
\end{scriptsize}
\end{table}

\begin{table}[h]
\caption[]{Performance evaluation of MIT-BIH Arrhythmia data (assume the client request prediction for a total of ten ECG signals)}\label{tab_d2}
\begin{scriptsize}
\begin{center}
\begin{tabular}{c c c c c c}
\toprule
& \multicolumn{5}{c}{No. of principal components}  \\
\cmidrule{2-6} 
 & 10 & 20 & 40 & 60 & 100 \\
\midrule
 \begin{tabular}{@{}c@{}} AUC \\ (proposed scheme) \end{tabular} & 98.39 & 98.78 & 98.88 & 98.78 & 98.66 \\
  \begin{tabular}{@{}c@{}} AUC \\ (non-secure SVM) \end{tabular} & 98.39 & 98.78 & 98.88 & 98.78 & 98.66 \\
Computation & 10 ms & 30 ms & 30 ms & 50 ms & 100 ms \\  
Communication* & 3.3 Mb & 6.7 Mb & 13.4 Mb & 20.1 Mb & 33.5 Mb \\  
\bottomrule
\end{tabular}
\end{center}
~~~~~~* 32-bit precision.
\end{scriptsize}
\end{table}

\begin{figure}[h]
\centering
\includegraphics[scale=0.5]{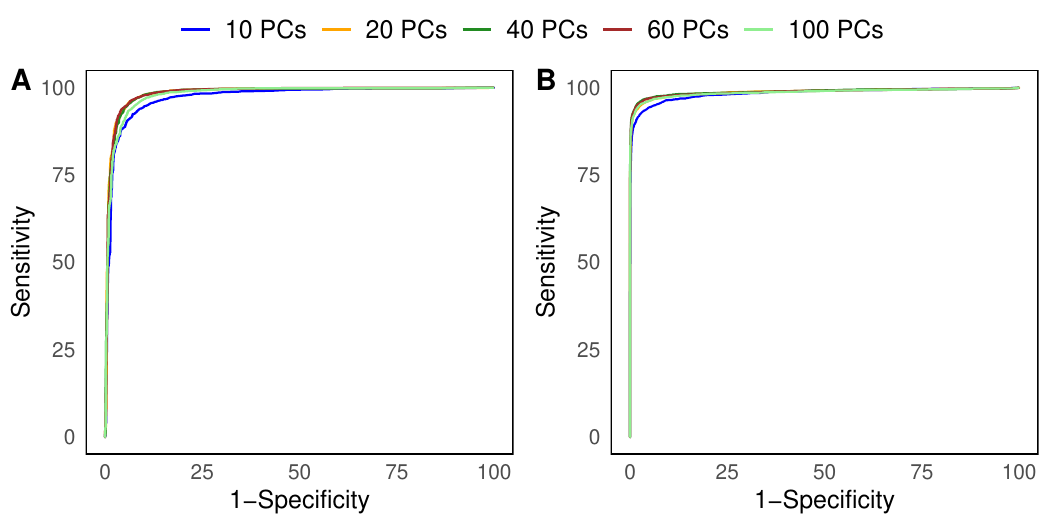}
\caption{ROC curve for PTB data (A) and MIT-BIH data (B) with different number of principal components (PCs). }\label{roc}
\end{figure}

The prediction accuracy, computation and communication cost for different numbers of principal components are presented in Table \ref{tab_d1} (PTB data) and \ref{tab_d2} (MIT-BIH data). SVM with radial kernel is used for model prediction. The proposed scheme has no accuracy degradation compared to non-secure SVM. The SVM prediction accuracy varies slightly for different dimensions of new feature space (i.e., No. of principal components) (Figure \ref{roc}). The model using smaller number of principal components requires lower computation and communication cost. Reducing the number of principal components does not significantly reduce model AUC but largely reduces the computation and communication cost.


\section{Conclusions}
In this paper, we propose a real-time privacy preserving disease prediction scheme for ECG data with no accuracy degradation. Experiments demonstrate the high computation efficiency and accuracy of the proposed scheme. Using a device with 4Gb RAMs, it takes less than 1 second to perform privacy preserving disease diagnosis using a diagnosis model trained from over 80000 samples. Future work includes investigating more efficient scheme to further decrease the communication cost.

\section*{Acknowledgments}
This work was supported by the National Institutes of Health [R01 GM118737].



\appendices

\section{Feature extraction algorithms}\label{appendix_LDA ICA}
{\bf LDA.} Linear Discriminant Analysis (LDA) provides the highest possible discrimination among different classes in the data. Suppose $S_W$ is the within class covariance matrix and $S_B$ is the between class covariance matrix. The projection matrix is defined as 
\begin{displaymath}
W_{lda}=\text{arg }\underset{w}{\text{max}}\frac{|W^TS_BW|}{|W^TS_WW|}.
\end{displaymath}
The dimension of the signal reduces from $q$ to $p$ by
\begin{displaymath}
z_i=W_{lda}^T x_i ~~~~~~~i=1,2,\cdots,n.
\end{displaymath}

{\bf ICA.} Independent component analysis (ICA) assumes that $X$ is linearly mixed with source signals such that 
\begin{displaymath}
X=AS
\end{displaymath}
where $A\in \mathbb{R}^{q\times q}$ is the weight matrix and $S$ is the matrix composed of source signals.

ICA reduces the dimension of $x_i$ from $q$ to $p$ as follows.
\begin{displaymath}
z_i=W_{ica}^Tx_i
\end{displaymath}
where $W_{ica}\in \mathbb{R}^{q\times p}$ is the projection matrix. $W_{ica}^T$ equals the first $p$ rows of $A^{-1}$. 

The procedure to derive $W_{ica}$ was given in \cite{ICA1,signal_2010,signal_2013}.

\section{Commutative encryption matrix}\label{appendixA}

A polynomial function $f(v)$ annihilates matrix $B$ if $f(B)=0$ (Section 3.3 in \cite{matrix}). The minimal polynomial of matrix $B$, $f_B(v)$, is the monic polynomial of minimum degree that annihilates $B$. $f_B(v)$ is unique and $f_B(\lambda)=0$ if and only if $\lambda$ is an eigenvalue of matrix $B$. Given any monic polynomial $f(v)$ and matrix $B$, $f(B)=0$ if and only if there is a monic polynomial $h(v)$ such that $f(v)=h(v)f_B(v)$ ($f_B(v)$ is the minimal polynomial of matrix $B$). So for any monic polynomial $f(v)$ with $f(B)=0$, each eigenvalue of $B$ is $f(v)$'s root.

To prove that $B_{ii}$ is determined by the random coefficients $(b_{i1},b_{i2},\cdots,b_{iq})$ ($i=1,2$) of matrix polynomials, we define three polynomial functions, $f_{1}(v)=\underset{j=1}{\overset{q}{\sum}}b_{1j}v^j$, $f_{2}(v)=\underset{j=1}{\overset{q}{\sum}}b_{2j}v^j$ and $f(v)=f_1(v)-f_2(v)$. It is easy to verify that $B_{11}=f_{1}(B_0)$ and $B_{22}=f_{2}(B_0)$. So $B_{11}=B_{22}$ is equivalent to $f(B_0)=0$. Using the above described principle, $B_{11}=B_{22}$ if and only if each eigenvalue of $B_0$ is a root of $f(v)$. Let $\lambda_i$ $(i=1,\cdots,q)$ be the $q$ unique eigenvalues of $B_0$. Suppose all these eigenvalues are the roots of $f(v)$. It can also be expressed as $\left( \begin{array}{cccc} \lambda_1&\lambda_1^2&\cdots&\lambda_1^{q} \\ \lambda_2&\lambda_2^2&\cdots&\lambda_2^{q} \\ \vdots&\vdots& \ddots &\vdots \\ \lambda_{q}&\lambda_{q}^2&\cdots&\lambda_{q}^{q} \end{array} \right)\left( \begin{array}{c} b_{11} \\ b_{12} \\ \vdots \\ b_{1q} \end{array} \right)=\left( \begin{array}{cccc} \lambda_1&\lambda_1^2&\cdots&\lambda_1^{q} \\ \lambda_2&\lambda_2^2&\cdots&\lambda_2^{q} \\ \vdots&\vdots& \ddots &\vdots \\ \lambda_{q}&\lambda_{q}^2&\cdots&\lambda_{q}^{q} \end{array} \right)\left( \begin{array}{c} b_{21} \\ b_{22} \\ \vdots \\ b_{2q} \end{array} \right)$. For ease of interpretation, let the $q\times q$ dimensional matrix shown in both sides of the equation be $\Lambda$, $b^{(1)}=\left( \begin{array}{c} b_{11} \\ b_{12} \\ \vdots \\ b_{1q} \end{array} \right)$, and $b^{(2)}=\left( \begin{array}{c} b_{21} \\ b_{22} \\ \vdots \\ b_{2q} \end{array} \right)$. The equation can be expressed as $\Lambda b^{(1)}=\Lambda b^{(2)}$. As a Vandermonde matrix with each $\lambda_i\ne \lambda_j$ ($i\ne j$), the rank of $\Lambda$ is $q$, i.e., $rank(\Lambda)=q$. Because $rank(\Lambda)=rank(\Lambda,\Lambda b^{(1)})=q$, there is only one solution for $b^{(2)}$ given $\Lambda$ and $\Lambda b^{(1)}$. So $B_{11}=B_{22}$ if and only if $b^{(1)}=b^{(2)}$ which indicates that $B_{ii}$ is determined by $b^{(i)}$ ($i=1,2$).

\section{Privacy evaluation}\label{appendix_plot}
\begin{figure}[h]
\centering
\includegraphics[scale=0.5]{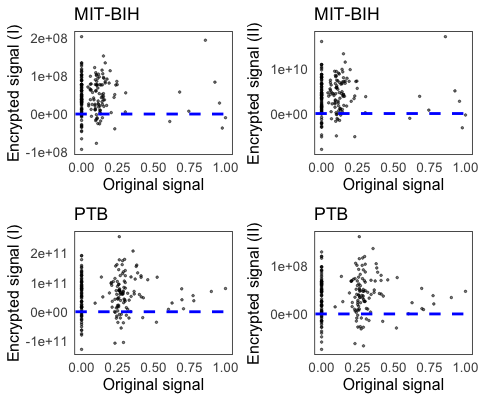}
\caption{Scatterplot showing the difference of the original signal ($t$), the encrypted signal I ($B_{11}t$) and II ($B_{22}t$). }\label{scatter_p1}
\end{figure}

\begin{figure}[h]
\centering
\includegraphics[scale=0.53]{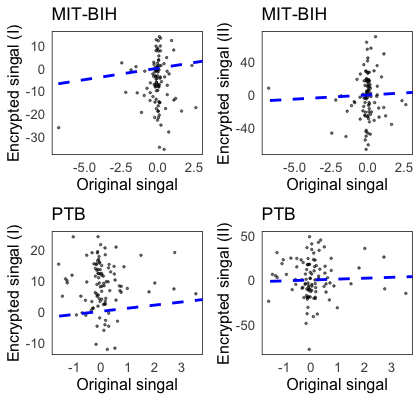}
\caption{Scatterplot showing the difference of the original signal in the new feature space ($Wt$), the encrypted signal I ($A_1Wt$) and II ($A_2Wt$). }\label{scatter_p2}
\end{figure}

\bibliographystyle{IEEEtran}
\bibliography{references}

\end{document}